\documentclass[review,authoryear]{elsarticle}

\usepackage[utf8]{inputenc}
\usepackage{longtable,booktabs}
\usepackage{threeparttable} 
\usepackage{threeparttablex}
\usepackage{dcolumn}
\usepackage[margin=3cm]{geometry}
\usepackage{lineno,hyperref}
\usepackage{subfig}
\usepackage{fancyhdr}
\usepackage{afterpage}
\usepackage{graphicx}

\usepackage{rotating, float, caption}
\modulolinenumbers[5]

\graphicspath{{FiguresDraft/}}

\usepackage{calc} 
\usepackage{amsmath,amssymb,latexsym} 
\usepackage{enumerate} 
\usepackage[dvipsnames]{xcolor}
\usepackage{amsmath,amssymb,latexsym} 
\usepackage{enumitem} 

\biboptions{numbers,sort&compress}

\journal{Journal of Statistical Planning and Inference}

\DeclareMathOperator{\PP}{\mathbf{P}}

 \DeclareMathOperator{\Exp}{Exp}
\DeclareMathOperator{\Gam}{Ga} 
\DeclareMathOperator{\Dir}{Dir} 
\DeclareMathOperator{\Bern}{Bern}
\DeclareMathOperator{\Unif}{Unif}

\begin{document}

\begin{frontmatter}

\title{Bayesian analysis of ranking data with the\\constrained Extended Plackett-Luce model}


\author[mymainaddress]{Cristina Mollica\corref{mycorrespondingauthor}}
\cortext[mycorrespondingauthor]{Corresponding author}
\ead{cristina.mollica@uniroma1.it}

\author[mysecondaryaddress]{Luca Tardella}

\address[mymainaddress]{Dipartimento di Metodi e Modelli per l'Economia, il Territorio e la
  Finanza, Sapienza Universit\`{a} di Roma}
\address[mysecondaryaddress]{Dipartimento di Scienze Statistiche, Sapienza Universit\`{a} di Roma}

\begin{abstract}
Multistage ranking models, including the popular Plackett-Luce distribution (PL), rely on the assumption that the ranking process is performed sequentially, by assigning the positions from the top to the bottom one (\textit{forward order}). A recent contribution to the ranking literature relaxed this assumption with the addition of the discrete-valued \textit{reference order} parameter, yielding the novel \textit{Extended Plackett-Luce model} (EPL). Inference on the EPL and its generalization into a finite mixture framework was originally addressed from the frequentist perspective. 
In this work, we propose the Bayesian estimation of the EPL with order constraints on the reference order parameter. 
The proposed restrictions reflect a meaningful rank assignment process.
By combining the restrictions with the data augmentation strategy and the conjugacy of 
the Gamma prior distribution with the EPL, we 
facilitate the construction
of a tuned joint Metropolis-Hastings algorithm within Gibbs sampling
to simulate from the posterior distribution. 
The Bayesian approach allows to address more efficiently the inference on the additional discrete-valued parameter and the assessment of its estimation uncertainty.
The usefulness of the proposal is illustrated with applications to simulated and real datasets.

\end{abstract}

\begin{keyword}
Ranking data \sep Plackett-Luce model \sep order constraints \sep Bayesian inference \sep Data augmentation \sep Gibbs sampling \sep Metropolis-Hastings
\end{keyword}

\end{frontmatter}


\section{Introduction}

\label{s:intro}

Ranking data are common in those experiments aimed at exploring preferences, attitudes or, more generically, choice behavior of a given population towards a set of \textit{items} or \textit{alternatives}~\citep{Vitelli,Gormley:Murphy-Royal,Yu2005,Vigneau1999}. A similar evidence emerges also in the sport context, yielding an ordering of the competitors, for instance players or teams, in terms of their ability or strength, see
~\cite{Henery-Royal},~\cite{Stern1990} and~\cite{Caron:Doucet}.

Formally, a \textit{ranking} $\pi=(\pi(1),\dots,\pi(K))$ of $K$ items is a sequence where the entry $\pi(i)$ indicates the rank attributed to the $i$-th alternative. Data can be equivalently collected in the ordering format $\pi^{-1}=(\pi^{-1}(1),\dots,\pi^{-1}(K))$, such that the generic component $\pi^{-1}(j)$ denotes the item ranked in the $j$-th position. Regardless of the adopted format, ranked observations are multivariate and, specifically,
correspond to permutations of the first $K$ integers. 

The statistical literature concerning ranked data modeling and analysis is broadly reviewed in \cite{Marden} and, more recently, in \cite{Alvo}. Several parametric distributions on the set of permutations $\mathcal{S}_K$ have been developed and applied to real data. A popular parametric family is the \textit{Plackett-Luce} model (PL), belonging to the class of the so-called \textit{stagewise ranking models}. The basic idea is the decomposition of the ranking process into $K-1$ stages, concerning the attribution of each position according to the \textit{forward order}, that is, the ordering of the alternatives proceeds sequentially from the most-liked to the least-liked item. The implicit forward order assumption has been released by~\cite{Mollica:Tardella} in the \textit{Extended Plackett-Luce model} (EPL). This relaxation allows for a more flexible dependence structure, hence for a better and possibly more parsimonious fitting of the observed ranking data. The PL extension in~\cite{Mollica:Tardella} relies on the introduction of the discrete \textit{reference order} parameter, indicating the rank assignment order, and its estimation was originally considered from the frequentist perspective. However, in that work there was no specific attention to the inferential ability to recover this extra parameter, as well as quantifying the estimation uncertainty. Indeed, the new EPL parameter space is of mixed-type (continuous and discrete) and, as far as the discrete component is concerned, there was little guidance 
on how to move efficiently towards the global optimal solution and only a multiple starting point approach was conceived. This resulted in a substantial computational burden with increasing $K$, which makes it less feasible a bootstrap approach to investigate parameter estimation uncertainty.

In this work, we 
investigate
a restricted version of the EPL with order constraints for the reference order parameter and 
detail
an original MCMC method to perform Bayesian inference. In particular, the considered parameter constraints formalize a meaningful
rank attribution process and we show how they can facilitate the definition of a joint proposal distribution for the Metropolis-Hastings (MH) step. 

The outline of the article is the following. 
After a review of
the main features of the EPL and the related data augmentation approach with latent variables,
the novel Bayesian EPL with order constraints is introduced in Section~\ref{s:pl}.
The detailed description of the MCMC algorithm to perform approximate posterior inference is presented in Section~\ref{s:MCMC}, whereas 
illustrative applications
to both simulated and
real ranking data follow in Section~\ref{s:appl}.
Final remarks and hints for future research are discussed in Section~\ref{s:conc}.

\section{The Bayesian Extended Plackett-Luce model}
\label{s:pl}


\subsection{Model specification}
\label{ss:modspec}
%

The PL
was introduced by~\cite{Luce} and~\cite{Plackett} and has a long history in the ranking literature for its numerous successful applications as well as for still inspiring new research developments.
The PL is a parametric class of ranking distributions indexed by the
\textit{support parameters} $\underline{p}=(p_1,\dots,p_K)$,
representing positive measures of liking for each item:
the higher the value of the support parameter $p_i$,
the greater the probability for the $i$-th item to be preferred at each selection stage. 
The expression of the PL distribution is
\begin{equation*}
\label{e:pl}
\PP_\text{PL}(\pi^{-1}|\underline{p})
=\prod_{t=1}^K\frac{p_{\pi^{-1}(t)}}{\sum_{v=t}^Kp_{\pi^{-1}(v)}}\qquad\pi^{-1}\in\mathcal{S}_K,
\end{equation*}
revealing the analogy of the underlying ranking 
selection
process with the sampling without replacement of the alternatives in order of preference. 

\begin{figure}
\begin{center}
\includegraphics[height=6cm, width=6cm]{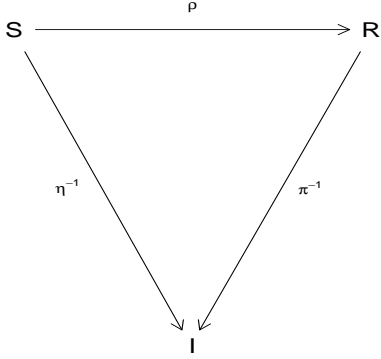}
\caption
{Mappings between the set of items, ranks and stages.}
\label{fig:mappings}
\end{center}
\end{figure}

The implicit assumption in the PL scheme is the forward ranking order, meaning that at the first stage the ranker reveals the item in the first position (most-liked alternative), at the second stage she assigns the second position and so on up to the last rank (least-liked alternative). \cite{Mollica:Tardella} suggested the extension of the PL by relaxing the canonical forward order assumption, in order to explore alternative meaningful ranking orders for the choice process and to increase the flexibility of the PL parametric family. Their proposal was realized by indexing the ranking order with an additional
model parameter  $\rho=(\rho(1),\dotsc,\rho(K))$, 
called \textit{reference order} and defined as the bijection between the stage set $S=\{1,\dots,K\}$ and the rank set $R=\{1,\dots,K\}$
\begin{equation*}
\rho:S\to R,
\end{equation*}
such that the entry $\rho(t)$ indicates the rank attributed at the $t$-th stage of the ranking process. 
Thus, $\rho$ is
a
discrete 
parameter given by a permutation of the first $K$ integers
and the composition $\eta^{-1}=\pi^{-1}\rho$ of an ordering
with a reference order 
yields the bijection between the stage set $S$ and the item set $I=\{1,\dots,K\}$
\begin{equation*}
\label{e:eta}
\eta^{-1}:S\to I.
\end{equation*}
The sequence $\eta^{-1}=(\eta^{-1}(1),\dotsc,\eta^{-1}(K))$ lists the items in order of selection, such that the component $\eta^{-1}(t)=\pi^{-1}(\rho(t))$ corresponds to the item chosen at stage $t$ and receiving rank $\rho(t)$. Figure~\ref{fig:mappings} summarizes the possible sequences defined as bijective mappings between the set of items, ranks and stages. 
 The probability of a generic ordering under the EPL can be written as
\begin{equation}
\label{e:EPL}
\PP_\text{EPL}(\pi^{-1}|\rho,\underline{p})=\PP_\text{PL}(\pi^{-1}\rho|\underline{p})
=\prod_{t=1}^K\frac{p_{\pi^{-1}(\rho(t))}}{\sum_{v=t}^Kp_{\pi^{-1}(\rho(v))}}\qquad\pi^{-1}\in\mathcal{S}_K.
\end{equation}
%
%
Hereinafter, we will shortly refer to 
the probability distribution in~\eqref{e:EPL} as $\text{EPL}(\rho,\underline{p})$. The quantities $p_i$'s
are still proportional to the probabilities for each item to be selected at the first stage,
but to be ranked
in the position
indicated by the first entry of $\rho$. 
Hence, we note that only in those extreme cases where either $\rho(1)=1$ or $\rho(1)=K$, one has a straightforward and natural interpretation of $\underline{p}$ as measures of liking or disliking respectively of each item. Obviously, this holds true for the standard PL as a special instance of the EPL with $\rho=\rho_\text{F}=(1,2,\dots,K)$, i.e., the \textit{identity permutation} also named \textit{forward order}. Similarly, this happens for the backward PL as a special case with $\rho=\rho_\text{B}=(K,K-1,\dots,1)=(K+1)-\rho_\text{F}$, i.e., the \textit{backward order}.

As in \cite{Mollica:Tardella2017}, the data  augmentation with
the latent quantitative variables $\underline{y}=(y_{st})$ for $s=1,\dots,N$ and $t=1,\dots,K$ crucially contributes to make the Bayesian inference for the EPL analytically tractable. Let $\underline\pi^{-1}=\{\pi_s^{-1}\}_{s=1}^N$ be the observed sample of $N$ orderings. The complete-data model can be specified as follows
%
%
\begin{eqnarray*}
\pi_s^{-1}|\rho,\underline{p} & \overset{\text{iid}}{\sim} & \text{EPL}(\rho,\underline{p})\qquad\qquad\quad\qquad s=1,\dots,N,\\
y_{st}|\pi_s^{-1},\rho,\underline{p} & \overset{\text{i}}{\sim} & \Exp\left(\sum_{\nu=t}^{K}p_{\pi_s^{-1}(\rho(\nu))}\right)\qquad t=1,\dots,K,
\end{eqnarray*}
where the auxiliary variables $y_{st}$'s are assumed to be conditionally independent on each other and exponentially distributed with rate parameter equal to the normalization term of the EPL. The complete-data likelihood turns out to be
%
\begin{equation}
\label{e:complik}
L_c(\rho,\underline{p},\underline{y})=\prod_{i=1}^Kp_i^Ne^{-p_i\sum_{s=1}^N\sum_{t=1}^Ky_{st}\delta_{sti}},
\end{equation}
where
%
\begin{equation*}
\delta_{sti}=\begin{cases}
      1\quad\text{ if }i\in\{\pi_s^{-1}(\rho(t)),\dots,\pi_s^{-1}(\rho(K))\},\\
      0\quad\text{ otherwise},
\end{cases}
\end{equation*}
with $\delta_{s1i}=1$ for all $s=1,\dots,N$ and $i=1,\dots,K$.


\subsection{Order constraints and prior distribution}
\label{ss:prior}
For the prior specification, we consider independence of $\underline{p}$ and $\rho$ and the following distributions
\begin{eqnarray*}
p_i & \overset{\text{i}}{\sim} & \Gam(c,d)\qquad i=1,\dots,K,\\
\rho & \sim & \Unif\left\{\tilde{\mathcal{S}}_K\right\}.
\end{eqnarray*}
The adoption of independent Gamma densities for the support parameters is motivated by the conjugacy with the model, as apparent by checking the form of the likelihood \eqref{e:complik}. In our analysis, we considered the hyperparameter setting $c=d=1$.
Differently from \cite{Mollica:Tardella}, we 
focus on a restriction $\tilde{\mathcal{S}}_K$ of the whole permutation space $S_K$ for the generation of the reference order, defined through the introduction of order constraints on the discrete parameter. Our choice can be motivated not only from a computational perspective, as widely illustrated 
in the next section, but also by the fact that in a preference elicitation process, not all the possible $K!$ orders seem to be equally natural, hence plausible.
Often the ranker has a clearer perception about her extreme preferences (most-liked and least-liked items), rather than middle positions. In this perspective,
the rank attribution process can be regarded as the result of a sequential ``top-or-bottom'' selection of the positions. At each stage, the ranker specifies either her best or worst choice among the available positions at that given step.
\begin{figure}
\begin{center}
\includegraphics[height=7cm, width=10cm]{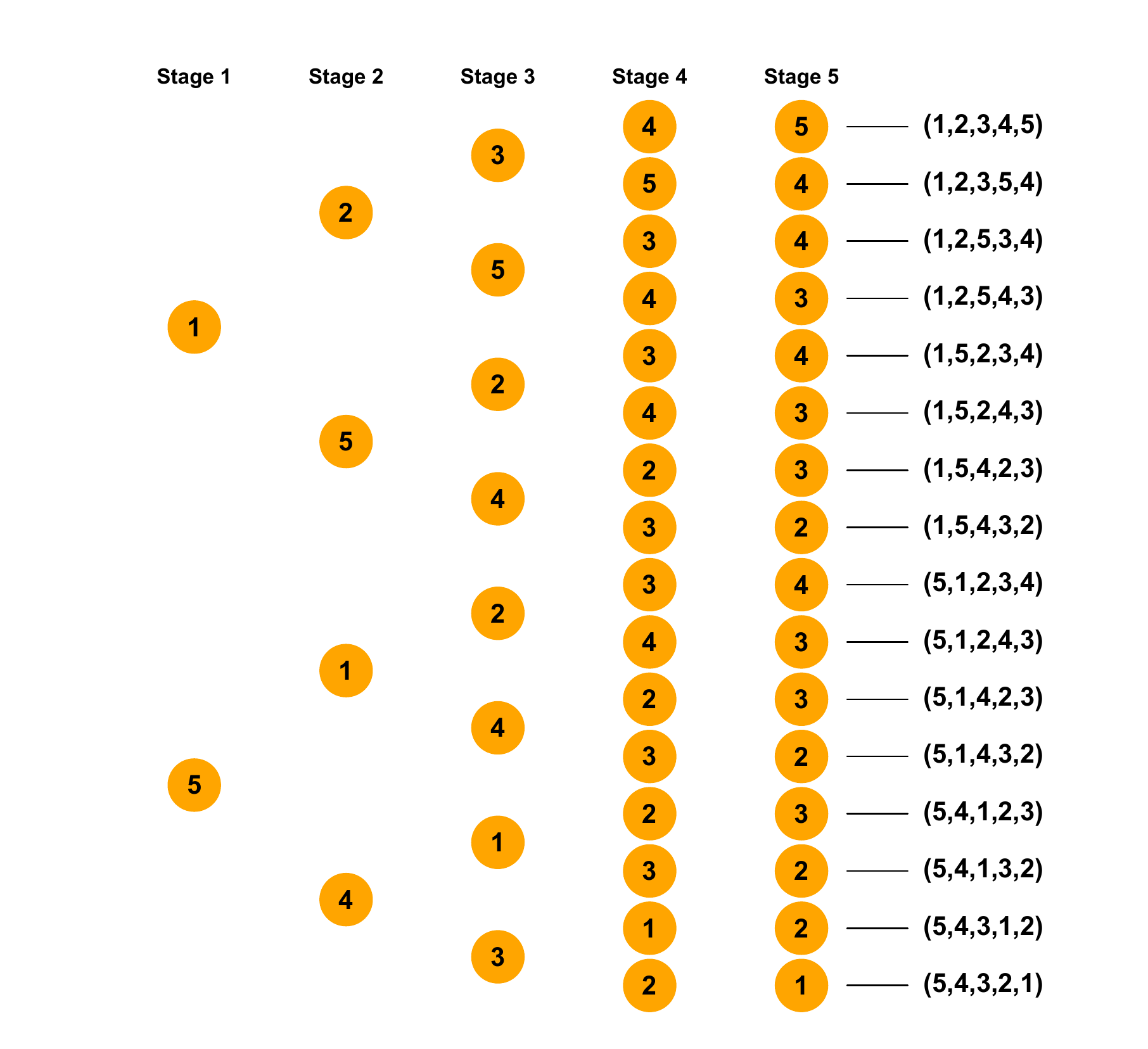}
\caption
{Restricted permutation space $\tilde S_5$ for the reference order parameter $\rho$ in the case of $K=5$ items.}
\label{fig:permtree}
\end{center}
\end{figure}
%
Figure~\ref{fig:permtree} shows the restricted permutation space $\tilde S_5$ for the reference order in the case of $K=5$ items. The first entry of the reference order, indicating the position assigned at the first stage, can be either $\rho(1)=1$ (most-liked item) or $\rho(1)=5$ (least-liked item). Let us suppose that 
at the first stage the ranker has ranked the item in the last position, i.e., $\rho(1)=5$; 
at the second stage, the ranker can express only her best or worst choice, conditionally on the fact that the last position has been already occupied; this means that either $\rho(2)=1$ or $\rho(2)=4$, and so on up to the final stage where the last component $\rho(K)$ is automatically determined. 

With this scheme, the reference order can be equivalently represented as a binary sequence $\underline{W}=(W_1,\dots,W_K)$,
where the generic $W_t$ component indicates whether the ranker makes a top or bottom decision at the $t$-th stage,
with the convention that $W_K=1$.
For the sake of notational compactness, one can formalize the mapping from the restricted permutation $\rho$ to  $\underline{W}$ 
with the help of a vector of non negative integers $\underline{F}=(F_1,\dots,F_K)$, 
where $F_t$ represents the number of top positions assigned before stage $t$.
In fact, by starting from positing by construction $F_1=0$, one can derive sequentially
\begin{equation*}
W_t=I_{[\rho(t)=\rho_\text{F}(F_t+1)]}
=\begin{cases}
      1\qquad \text{at stage $t$ the top preference is specified}, \\
      0\qquad \text{at stage $t$ the bottom preference is specified},
\end{cases}
\end{equation*}
where $I_{[E]}$ is the binary indicator of the event $E$ and  
$F_t=\sum_{\nu=1}^{t-1}W_\nu$ for $t=2,...,K$.
Note that, since the forward and backward orders $(\rho_\text{F}, \rho_\text{B})$ 
can be regarded as the two extreme benchmarks in the sequential construction of $\rho$,
this allows us to understand that 
$\rho_F(F_t+1)$ corresponds to the top 
position available at stage $t$.
Conversely, $B_t=(t-1)-F_t$
is the number of bottom positions assigned before stage $t$ and thus,
symmetrically, one can understand that $\rho_B(B_t+1)$ indicates the bottom
position available at stage $t$.
The inverse mapping from the binary vector $\underline{W}$ to the
constrained reference order $\rho \in \tilde{\mathcal{S}}_K$
can be written as follows
%
\begin{equation}
\label{ournotation}
\rho(t)=\rho_\text{F}(F_t+1)^{W_t}\rho_\text{B}(B_t+1)^{1-W_t}\qquad t=1,\dots,K.
\end{equation}
%
A toy example 
in the Appendix
can help to clarify
the notation adopted for the constrained reference order parameter.

From an inferential point of view, the order constraints of the ``top-or-bottom'' scheme are convenient for several reasons: i) as suggested by the binary representation of $\rho$, the size of $\tilde{\mathcal{S}}_K$ is equal to $2^{K-1}$, implying a reduction of the reference order space into a finite set with an exponential cardinality, rather than with factorial size as $\mathcal{S}_K$; ii) the restrictions lead to a more intuitive interpretation of the support parameters, since they become proportional to the probability for each item to be ranked either in the first or in the last position and iii) the order constraints facilitate the construction of a proposal distribution for the MH step to sample the reference order parameter, as better described in the next section.

\section{Bayesian estimation of the constrained EPL via MCMC}
\label{s:MCMC}
In this section, we describe an original MCMC algorithm to solve the Bayesian inference for the constrained EPL. 

We propose a tuned joint Metropolis-within-Gibbs sampling (TJM-within-GS) as simulation-based method to approximate the posterior distribution. Its distinguishing feature is the use of a suitably tuned MH algorithm relying on a joint proposal 
on the mixed-type parameter components $(\rho,\underline{p})$, 
combined with two other kernels which acts more specifically on the discrete component
$\rho$ and then on the continuous $(\underline{p},y)$.
In fact, although partitioning the parameter vector of the augmented space by using the components ($\rho,\underline{p},\underline{y}$) makes it possible to derive standard full-conditionals for a Gibbs sampling, this ends up being a 
difficult-to-implement and 
unsuccessful strategy in practice. 
On one hand,
the discrete full-conditional for $\rho$ involves a support with a rapidly-increasing cardinality (with $K$).
On the other hand the meaning of the support parameters is strictly related to the reference order
(mainly to its first component)
and this means that, 
when one keeps $\underline{p}$ fixed, the full-conditional of $\rho$ is very unlikely to 
move away from the current value, resulting in small local updates.
This argument inspired us a more successful alternative joint proposal strategy for a MH kernel.

\subsection{Tuned Joint Metropolis-Hastings step}
\label{ss:tjmh}
Let us denote with $\underline{\lambda}=(\lambda_1,\dots,\lambda_K)$ the vector of Bernoulli probabilities 
$$\lambda_t=\PP(W_t=1|W_1,\dots,W_{t-1})=\PP(\rho(t)=\rho_\text{F}(F_t+1)|\rho(1),\dots,\rho(t-1)).$$ 
A possible proposal distribution to be employed in the MH step for sampling the reference order
could have the following form
\begin{equation*}
\PP(\rho)=\prod_{t=1}^K\PP(\rho(t)|\rho(1),\dots,\rho(t-1))=\prod_{t=1}^K\lambda_t^{W_t}(1-\lambda_t)^{1-W_t}.
\end{equation*}
Nevertheless, preliminary implementations of a MH step of this kind
on synthetic data 
suggested that only a joint proposal distribution of the reference order and the support parameters allows for an adequate mixing of the resulting Markov Chain.
Thus, to simultaneously sample candidate values for $\rho$ and $\underline{p}$, we 
devised a Metropolis kernel $\mathcal{K}_{\text{TJM}}$ based on a
joint proposal distribution $g(\rho,\underline{p})$ with a specific decomposition of the dependence structure, given by
\begin{equation}
\label{e:jp}
g(\rho,\underline{p})=g(\rho(1))\times g(\underline{p}|\rho(1))\times \prod_{t=2}^Kg(\rho(t)|\underline{p},\rho(1),\dots,\rho(t-1)).
\end{equation}
The dependence structure in~\eqref{e:jp} shows that, after drawing the first component of $\rho$, the proposal 
can exploit
the sample evidence on the support parameters to guide the simulation of the remaining candidate entries of the reference order. In so doing, the generation of the two parameter vectors are linked to each other,
in order to mimic the target density and, hence, getting a better mixing chain.
Candidate values $(\tilde\rho,\tilde{\underline{p}})$
are jointly generated according to the following scheme:
\begin{enumerate}
\item sample the first component of $\rho$ (stage $t=1$)
$$\tilde W_1\sim\Bern(\tilde\lambda_1)\quad\Rightarrow\quad\tilde \rho(1)=\rho_\text{F}(1)^{\tilde W_1}\rho_\text{B}(1)^{1-\tilde W_1}=1^{\tilde W_1}K^{1-\tilde W_1}.$$
In our application, we set $\tilde \lambda_1=\PP(\tilde W_1=1)=0.5$;
\item sample the support parameters
$$\tilde{\underline{p}}|\tilde \rho(1)\sim\text{Dirich}(\alpha_0\times\underline{r}_{\tilde \rho(1)}),$$
where \text{Dirich} denotes the Dirichlet distribution, $\alpha_0$ is a scalar tuning parameter and $\underline{r}_{\tilde \rho(1)}$ is the vector collecting either the marginal top or bottom item relative frequencies
according to whether $\tilde\rho(1)=1$ or $\tilde\rho(1)=K$. 
Specifically, the $i$-th entry of $\underline{r}_{\tilde \rho(1)}$ is
$$r_{\tilde \rho(1)i}=\frac{1}{N}\sum_{s=1}^NI_{[\pi_s^{-1}(\tilde \rho(1))=i]}$$
and, in our analysis, we set $\alpha_0=50$;
\item sample the remaining entries of the reference order (from stage $t=2$ to stage $t=K-1$)
iteratively as follows: once selected the reference order component at stage $t-1$, we consider the two 
observed
contingency tables $\tilde\tau$ and $\tilde\beta$ having as first margin the item 
placed at the current reference order component $\tilde \rho(t-1)$ and as second margin, respectively, the item 
placed at the reference order component which can be possibly selected at the next stage, denoted as either $\rho_F(F_t+1)$
or $\rho_B(B_t+1)$.
The generic entries of the two contingency tables are
%
\begin{align*}
\tilde\tau_{ii't}&=\sum_{s=1}^NI_{[\pi_s^{-1}(\tilde \rho(t-1))=i,\pi_s^{-1}(\rho_F(F_t+1))=i']},\\
\tilde\beta_{ii't}&=\sum_{s=1}^NI_{[\pi_s^{-1}(\tilde \rho(t-1))=i,\pi_s^{-1}(\rho_B(B_t+1))=i']},
\end{align*}
%
%
corresponding to the actually observed joint frequencies counting how many times each item $i$ in the previous stage is followed by any other item $i'$ at the next stage. We then compare these frequencies with the corresponding expected frequencies $\tilde E_{ii't}$ under the EPL
by using a Monte Carlo approximation
\begin{align*}
\tilde\eta_s^{-1}(1),\dots,\tilde\eta_s^{-1}(t)|\tilde{\underline{p}}&\overset{\text{i}}{\sim}\text{PL}(\tilde{\underline{p}})\qquad s=1,\dots,N,\\
\tilde E_{ii't}&=\sum_{s=1}^NI_{[\tilde\eta_s^{-1}(t-1)=i,\tilde\eta_s^{-1}(t)=i']}
\end{align*}
%
and compute the following top and bottom distances
$$\tilde d_t^{T}=\sum_{i=1}^K\sum_{i'=1}^K(\tilde\tau_{ii't}-\tilde E_{ii't})^2\quad\text{and}\quad\tilde d_t^{B}=\sum_{i=1}^K\sum_{i'=1}^K(\tilde\beta_{ii't}-\tilde E_{ii't})^2.$$
The above distances are then suitably scaled
as follows
$$\tilde d_{t}=1-\frac{\tilde d_{t}^{T}}{\tilde d_{t}^{T}+\tilde d_{t}^{B}}$$
and exploited in order to mimic the conditional probability corresponding to the target distribution.
Indeed, we define the Bernoulli proposal probability of top selection at stage $t$ as
$$\tilde\lambda_{t}=\tilde d_{t}(1-2h)+h,$$
where $h\in(0,0.5)$ is a tuning parameter introduced to guarantee
a minimal positive probability $h$ for the bottom selection ($\tilde\lambda_{t} \geq h$). We set as default value $h=0.1$. Finally, for $t=2,\dots,K-1$, we sample
$$\tilde W_t\sim\Bern(\tilde\lambda_t)\quad\Rightarrow\quad\tilde \rho(t)=\rho_F(F_t+1)^{\tilde W_t}\rho_B(B_t+1)^{1-\tilde W_t}.$$
%
\end{enumerate}
The 
resulting
joint proposal probability of the candidate values
is
$$g(\tilde\rho,\tilde{\underline{p}})=\Dir(\tilde{\underline{p}}|\alpha_0\times\underline{r}_{\tilde \rho(1)})\prod_{t=1}^K\tilde\lambda_{t}^{\tilde W_t}(1-\tilde\lambda_{t})^{(1-\tilde W_t)}.$$
Hence, 
if we denote the observed-data likelihood with $L(\rho,\underline{p})$, the acceptance probability is equal to
\begin{equation*}
\alpha'=
\min\left\{\frac{g(\rho^{(l)},\underline{p}^{(l)})}{g(\tilde\rho,\tilde{\underline{p}})}
\frac{L(\tilde\rho,\tilde{\underline{p}})\prod_{i=1}^K\Gam(\tilde p_i|c,d)}{L(\rho^{(l)},\underline{p}^{(l)})\prod_{i=1}^K\Gam(p_i^{(l)}|c,d)},1\right\}
\end{equation*}
and the MH step ends with the classical acceptance/rejection of the candidate pair
%
\begin{equation*}
(\rho',\underline{p}')=\begin{cases}
      (\tilde\rho,\tilde{\underline{p}})\qquad \text{\qquad if $\log(u')<\log(\alpha')$}, \\
      (\rho^{(l)},\underline{p}^{(l)})\qquad \text{otherwise},
\end{cases}
\end{equation*}
where $u'\sim\Unif(0,1)$ and $(\rho^{(l)},\underline{p}^{(l)})$ is the current pair.

In order to facilitate mixing, we combine the just illustrated Metropolis kernel $\mathcal{K}_{\text{TJM}}$ by composing it with two additional kernels,
labelled as $\mathcal{K}_{\text{SM}}$ (Swap Move) and $\mathcal{K}_{\text{GS}}$ (Gibbs Sampling).
Hence, the MCMC simulation is based on the composition kernel 
$\mathcal{K} = \mathcal{K}_{\text{TJM}} \circ \mathcal{K}_{\text{SM}} \circ \mathcal{K}_{\text{GS}}$.
Indeed, the 
%
$\mathcal{K}_{\text{SM}}$  kernel focuses on local moves of the discrete component $\rho$, 
whereas 
$\mathcal{K}_{\text{GS}}$
aims at improving the mixing of the continuous component.
The 
kernel $\mathcal{K}_{\text{SM}}$ is illustrated in the next section,
while 
$\mathcal{K}_{\text{GS}}$ is just a full Gibbs sampling cycle involving the $(\underline{p},\underline{y})$ components and is detailed in Section~\ref{ss:tjmhgibbs}. 

\subsection{Swap move}
\label{ss:tjmh}
We remind that the values $(\rho',\underline{p}')$ have to be regarded as temporary parameter drawings. We decided to accelerate the exploration of the parameter space by including an intermediate kernel $\mathcal{K}_{\text{SM}}$ of the $\rho$ component only that attempts a possible local move w.r.t. to the current value. We label the possibly successful update of this kernel as \textit{Swap Move} (SM).
%
In fact, the idea relies on a random swap of two adjacent components of $\rho'$. Let $M\in\{1,\dots,K-1\}$ be the number of applicable contiguous swaps on $\rho'$, such that the order constraints of $\tilde{\mathcal{S}}_K$ still hold in the returning sequence, and $\{t_1,\dots,t_M\}$ be the indexes of the entries of $\rho'$ that can be switched with the consecutive ones. Note that the last two entries can be always swapped, meaning that $t_M=K-1$. The additional MH step consists in proposing a further reference order with a randomly selected SM. Specifically, one first simulate 
$$t^*\sim\Unif\{t_1,\dots,t_M\}$$
and then define the new candidate as
$$\rho''=(\rho'(1),\dots,\rho'(t^*+1),\rho'(t^*),\dots,,\rho'(K)).$$
Finally, by computing the acceptance probability as
\begin{equation*}
\alpha''=
\min\left\{\frac{g(\rho',\underline{p}')}{g(\rho'',\underline{p}')}
\frac{L(\rho'',\underline{p}')}
{L(\rho',\underline{p}')}
,1\right\},
\end{equation*}
the sampled value of the reference order at the $(l+1)$-th iteration turns out to be
\begin{equation*}
\rho^{(l+1)}=\begin{cases}
      \rho''\qquad \text{if $\log(u'')<\log(\alpha'')$}, \\
      \rho'\qquad \text{otherwise},
\end{cases}
\end{equation*}
where $u''\sim\Unif(0,1)$.

\subsection{Tuned Joint Metropolis-within-Gibbs-sampling}
\label{ss:tjmhgibbs}
At the generic iteration $(l+1)$, the TJM-within-GS
iteratively alternates the following simulation steps
\begin{eqnarray*}
\rho^{(l+1)},\underline{p}' & \sim & \text{TJM} \circ \text{SM},\\
y_{st}^{(l+1)}|\pi_s^{-1},\rho^{(l+1)},\underline{p}' & \sim & \Exp\left(\sum_{i=1}^K\delta_{sti}^{(l+1)}p_i'\right),\\
p_{i}^{(l+1)}|\underline\pi^{-1},\underline{y}^{(l+1)},\rho^{(l+1)} & \sim & \text{Ga}\left(c+N,d+\sum_{s=1}^{N}\sum_{t=1}^{K}\delta_{sti}^{(l+1)}y_{st}^{(l+1)}\right).
\end{eqnarray*}
The above outline shows that the full-conditional of the unobserved continuous variables $y$'s is given by construction of the complete-data model specified in Section~\ref{ss:modspec}, whereas the full-conditional of the support parameters is induced by the partial conjugate structure, requiring a straightforward update of the corresponding Gamma priors.

\section{Illustrative applications}
\label{s:appl}

\subsection{Simulated data}
\label{ss:appsim}
In order to verify the efficacy of our MCMC strategy, as well as the ensuing inferential ability of the proposed Bayesian framework, we have setup the following simulation plan: we considered a grid of simulation settings combining different number of items 
$K \in \{5,10,20 \}$ 
and sample sizes
$N \in \{50,200,1000,10000\}$. 
For each distinct pair $(K,N)$,
we replicated $100$ times the simulation of datasets from the 
constrained EPL model 
by varying the parameter configuration: for each simulated sample $\underline\pi^{-1}_{(R)}$ with $R=1,\dots,100$, we fixed a true reference order $\dot\rho^{(R)}$ and 
a true support parameter vector $\dot{\underline{p}}^{(R)}$ by drawing $\dot\rho^{(R)}$ uniformly in the restricted space $\tilde{\mathcal{S}}_K$ and  the components $\dot{p}_i^{(R)}$ i.i.d. from a uniform distribution. 

For each replication $R$, we run the TJM-within-GS described in Section~\ref{s:MCMC} for a total of 10000 iterations and 2000 were discarded as burn-in phase. The resulting simulations were considered as an approximation of the posterior distribution. 
The MCMC results were satisfactory in terms of convergence diagnostics and mixing and are omitted since they have basically the same qualitative behaviour displayed in the following real data applications. 
Hence, we limit ourselves to display and comment the results related specifically to the ability of the approximated marginal posterior distribution
to recover the known true reference order
by averaging the results over the 100 replicated datasets $\underline\pi^{-1}_{(R)}$. 
In fact, for each replication $R$, we have focussed only on the approximated marginal posterior distribution of $\rho$, denoted as $\pi(\rho|\underline\pi^{-1}_{(R)})$, and used the corresponding posterior mode $\hat{\rho}^{(R)}$ as the point estimate of the unknown reference order $\dot\rho^{(R)}$.
From Table~\ref{t:simul} we can appreciate how frequently the posterior mode $\hat{\rho}^{(R)}$ matches the true reference order $\dot\rho^{(R)}$ in terms of the percentage of true recoveries (\% recovered). We note that this percentage consistently grows with $N$ and, on average, a larger portion of posterior mass $\bar{\pi} ( \rho = \hat{\rho} | \underline\pi^{-1})$ is assigned to the matching mode.
Moreover, by considering all the replications, even those in which there is no match, the posterior mode ensures a consistently decreasing (with $N$) average relative Kendall distance between the true reference order and the estimated one, see the column $\bar{d}_{\text{Kend}}(\dot\rho,\hat{\rho})$ of Table~\ref{t:simul}. 
This means that the whole posterior distribution is consistently concentrating around the true $\dot\rho^{(R)}$.
Since the range of the Kendall distance depends on $K$, we rescaled it by $K(K-1)/n$ in order to have a relative index ranging over $[0,1]$ regardless of $K$. Finally, if we look at a fixed $N$, we observe that the inferential properties worsen for increasing values of $K$, as expected.

\begin{ThreePartTable}
\begin{TableNotes}
      \item[-] $\bar{d}_{\text{Kend}}(\dot\rho,\hat{\rho}) = \frac{1}{100} \sum_{R=1}^{100}  d_{\text{Kend}}(\dot\rho^{(R)},\hat{\rho}^{(R)})$
      \item[-] {\small $\bar{\pi} ( \rho = \hat{\rho} | \underline\pi^{-1})=\frac{1}{100} \sum_{R=1}^{100}  I_{[\dot\rho^{(R)}=\hat{\rho}^{(R)}]}\pi(\rho=\hat{\rho}^{(R)} |\underline\pi^{-1}_{(R)})$}
      \item[-] \% recovered = $ \sum_{R=1}^{100}  I_{[\dot\rho^{(R)}=\hat{\rho}^{(R)}]}$
\end{TableNotes}
\begin{longtable}[]{@{}lrcr@{}}
\caption{Results of the inference on the reference order for the simulated data.}
 \label{t:simul} \\
\toprule
$(K,N)$ & \(\bar{d}_{\text{Kend}}(\dot\rho,\hat{\rho})\) &
\(\bar{\pi} ( \rho = \hat{\rho} | \underline\pi^{-1})\) & \%
 recovered \tabularnewline
\midrule
\endhead
\endfoot
\insertTableNotes         
\endlastfoot
$(5,50)$ & 0.27 & 0.77 & 56\tabularnewline
$(5,200)$ & 0.10 & 0.94 & 84\tabularnewline
$(5,1000)$ & 0.00 & 0.98 & 100\tabularnewline
$(5,10000)$ & 0.00 & 1.00 & 100\tabularnewline
$(10,50)$ & 0.31 & 0.30 & 21\tabularnewline
$(10,200)$ & 0.19 & 0.69 & 49\tabularnewline
$(10,1000)$ & 0.07 & 0.89 & 81\tabularnewline
$(10,10000)$ & 0.00 & 1.00 & 100\tabularnewline
$(20,50)$ & 0.39 & 0.09 & 1\tabularnewline
$(20,200)$ & 0.30 & 0.09 & 4\tabularnewline
$(20,1000)$& 0.16 & 0.49 & 32\tabularnewline
$(20,10000)$ & 0.01 & 0.94 & 92\tabularnewline
\bottomrule
\end{longtable}
\end{ThreePartTable}

\subsection{Application to the LFPD data}
\label{ss:applfpd}
We applied our Bayesian constrained EPL to the Large Fragment Phage Display data (LFPD), come up from a recent technology of epitope mapping for breast cancer and involving multivariate quantitative measurements of the binding between human antibodies and $K=11$ partially overlapping fragments of the HER2 oncoprotein. The fragments were denoted sequentially with the labels Hum 1,$\dots$, Hum 11. For details on the biological foundation of the LFPD experiment, see~\citep{Gabrielli:al}. Since the observations were originally quantitative, we preliminary converted the binding profiles into ranked sequences, such that $\pi^{-1}(1)$ represents the site with the highest absorbance level. The ranked version of the entire LFPD dataset was previously analyzed by~\cite{Mollica:Tardella} with the estimation of a finite EPL mixture within the frequentist domain. Their ranking-based approach was proved to provide a more robust evidence for the characterization of the observed disease status.
%
\begin{figure}
\begin{center}
\includegraphics[width=12cm,height=8cm]{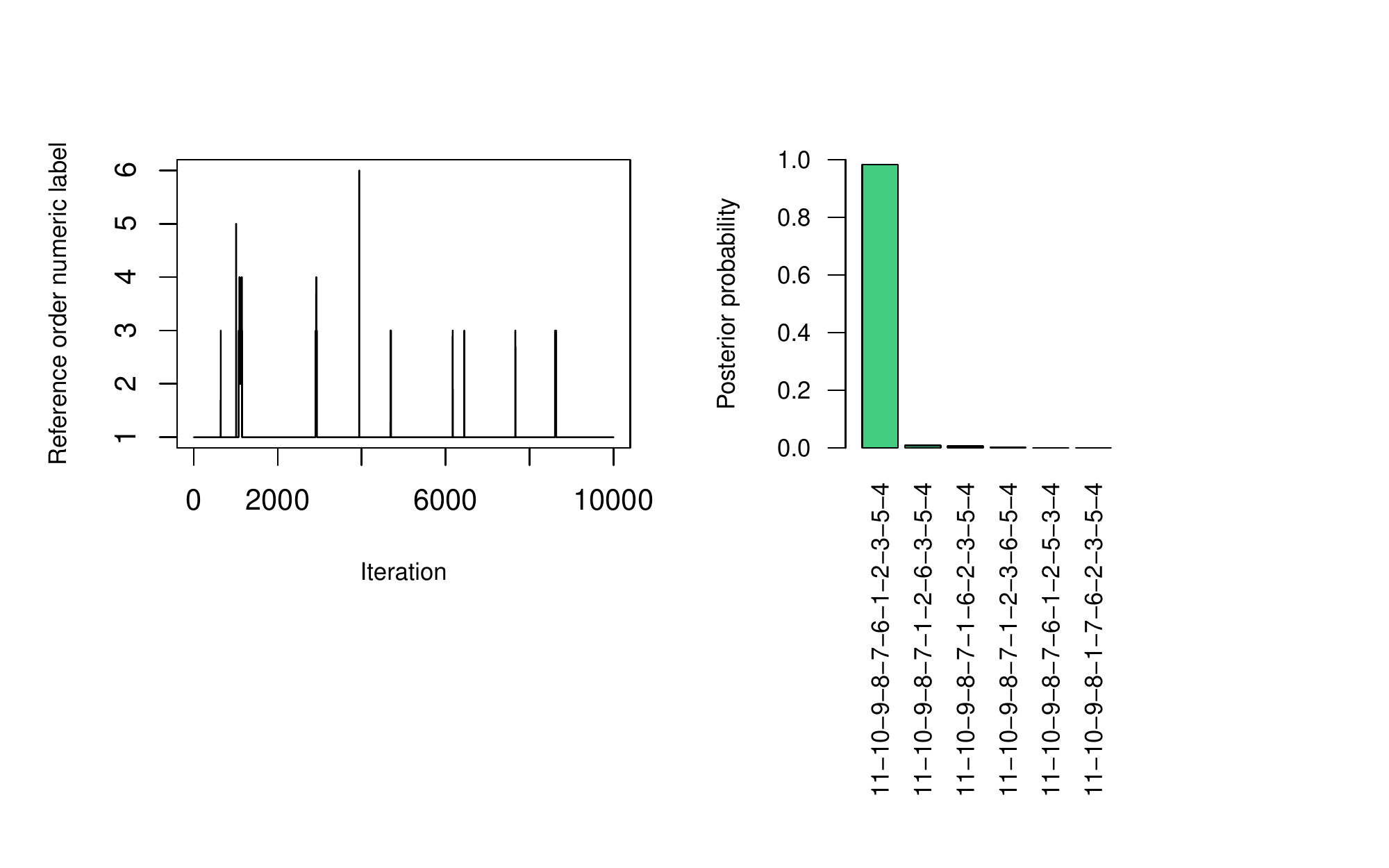}%
\caption{Traceplot (left) and top-6 posterior probabilities (right) for the reference order parameter.}
\label{fig:postrefMBC}
\end{center}
\end{figure}
The Bayesian restricted EPL
was fitted to 
the subsample of 19 patients diagnosed with metastastic breast cancer.
We run the TJM-within-GS for a total of 20000 iterations and 10000 were discarded as burn-in phase. The MCMC algorithm was launched with four random dispersed starting points to explore the mixing performance of the sampler over the parameter space. Indeed, the four chains were found to be consistent with respect to the initial values.
%
%
We show the results relative to the marginal posterior distribution on the reference order in Figure~\ref{fig:postrefMBC},
with the corresponding probability masses reported in Table~\ref{t:agreeEPL1}.
With a distinctive probability equal to 0.9832, the posterior modal reference order turns out to be $\hat\rho=(11,10, 9, 8, 7, 6, 1, 2, 3, 5, 4)$, that is, very different from the canonical forward order. The posterior means of the support parameters are shown in Table~\ref{t:agreeEPL2} and, combined with $\hat\rho$, yield an estimated modal ordering equal to $(2, 10, 7, 1, 11, 8, 6, 9, 3, 5, 4)$.
These results suggest that the epitope mapping, aimed at detecting protein fragments with higher absorbance level, could be combined with the identification of the protein sites with lower or absent binding for a better understanding of the role of the HER2 oncoprotein in the breast cancer diagnosis. 

%

%
\begin{table}[]
\caption{Inferential results for the subsample of the LFPD dataset.}
\label{t:agreeEPL}
\centering
\subfloat[][Top-5 posterior probabilities of $\rho$.] 
{\begin{tabular}{cc}
\hline
$\rho$ & $\pi(\rho|\underline\pi^{-1})$\\
\hline
(11,10,9,8,7,6,1,2,3,5,4) & .9832\\
(11,10,9,8,7,1,6,2,3,5,4) & .0080\\
(11,10,9,8,7,1,2,3,6,5,4) & .0060\\
(11,10,9,8,7,6,1,2,5,3,4) & .0025\\
(11,10,9,8,1,7,6,2,3,5,4) & .0002\\
\hline
\\
\end{tabular}\label{t:agreeEPL1}}\\
\subfloat[Posterior means of the support parameters.]
{\begin{tabular}{cccccccccccc}
\hline
Hum & 1 & 2 & 3 & 4 & 5 & 6 & 7 & 8 & 9 & 10 & 11  \\
\hline
$\hat{p}_i$ & .0070 &  .0596 &  .1613 &  .1886 &  .1836 &  .1015 &  .0287 &  .0771 &  .1492 &  .0343 &  .0092 \\
\hline
\end{tabular}\label{t:agreeEPL2}}
\end{table}
%


\subsection{Application to the sport data}
\label{ss:appsport}
%

\begin{sidewaysfigure}[!h]
\subfloat{\includegraphics[scale=0.25]{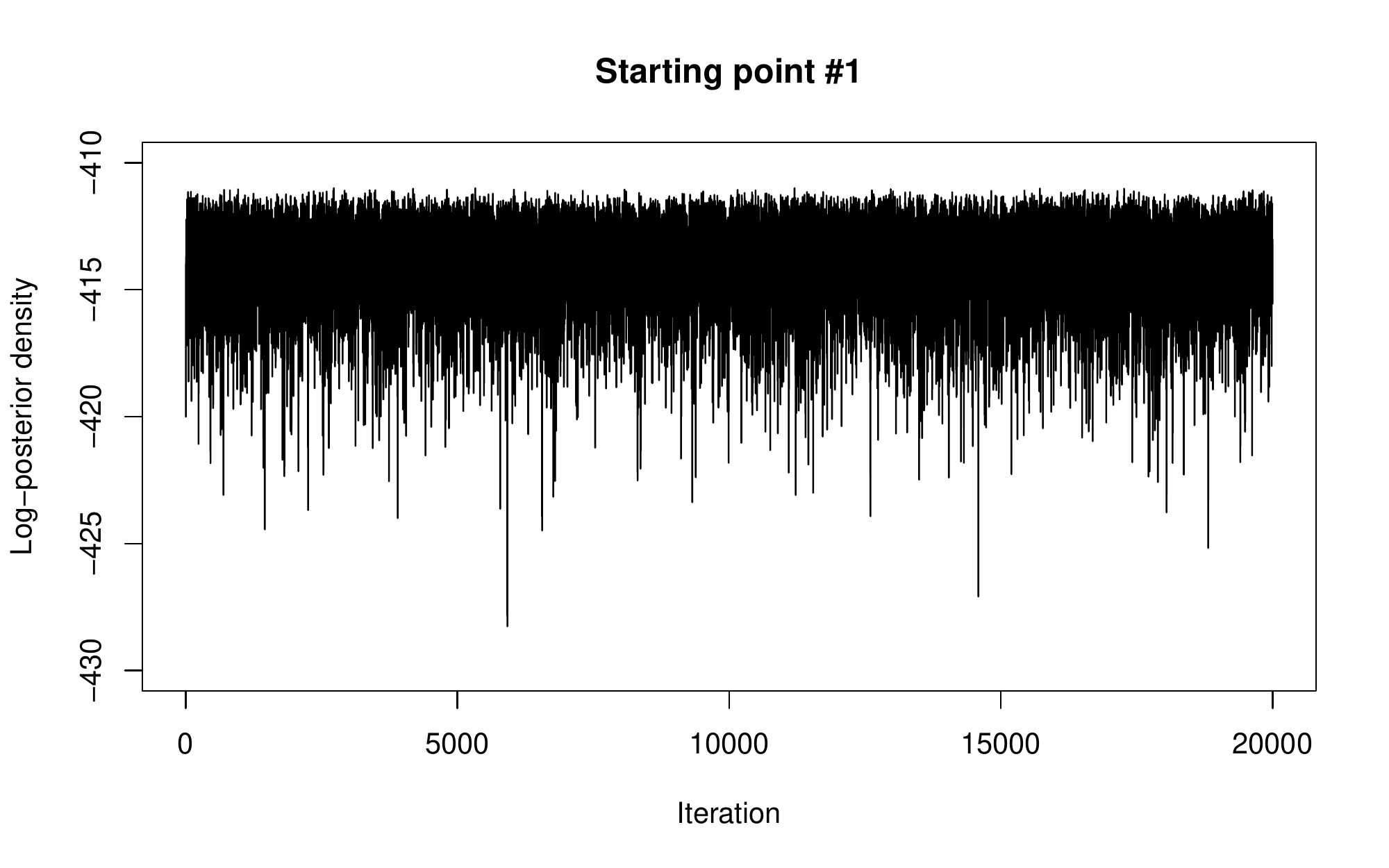}}
\subfloat{\includegraphics[scale=0.25]{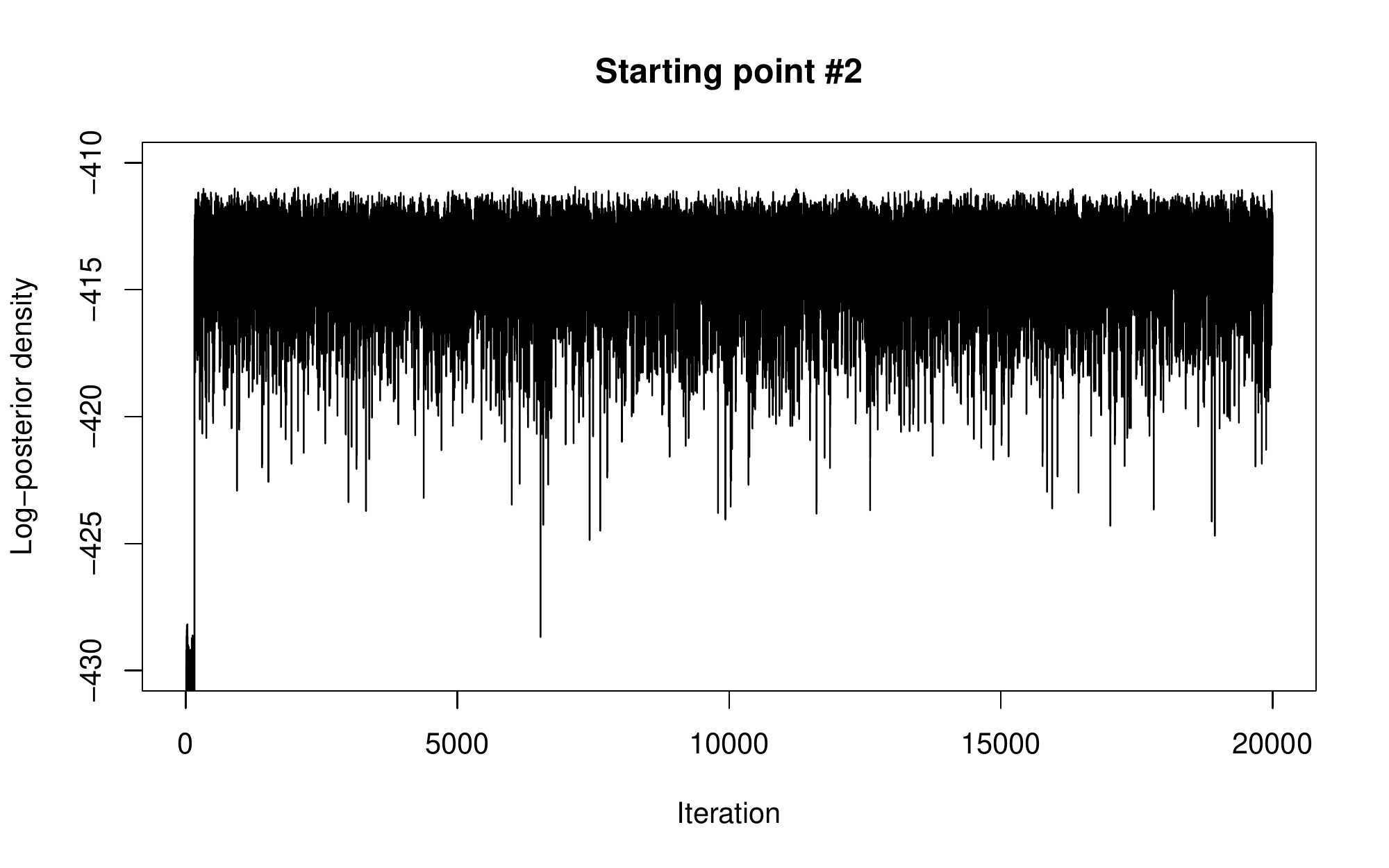}}
\subfloat{\includegraphics[scale=0.25]{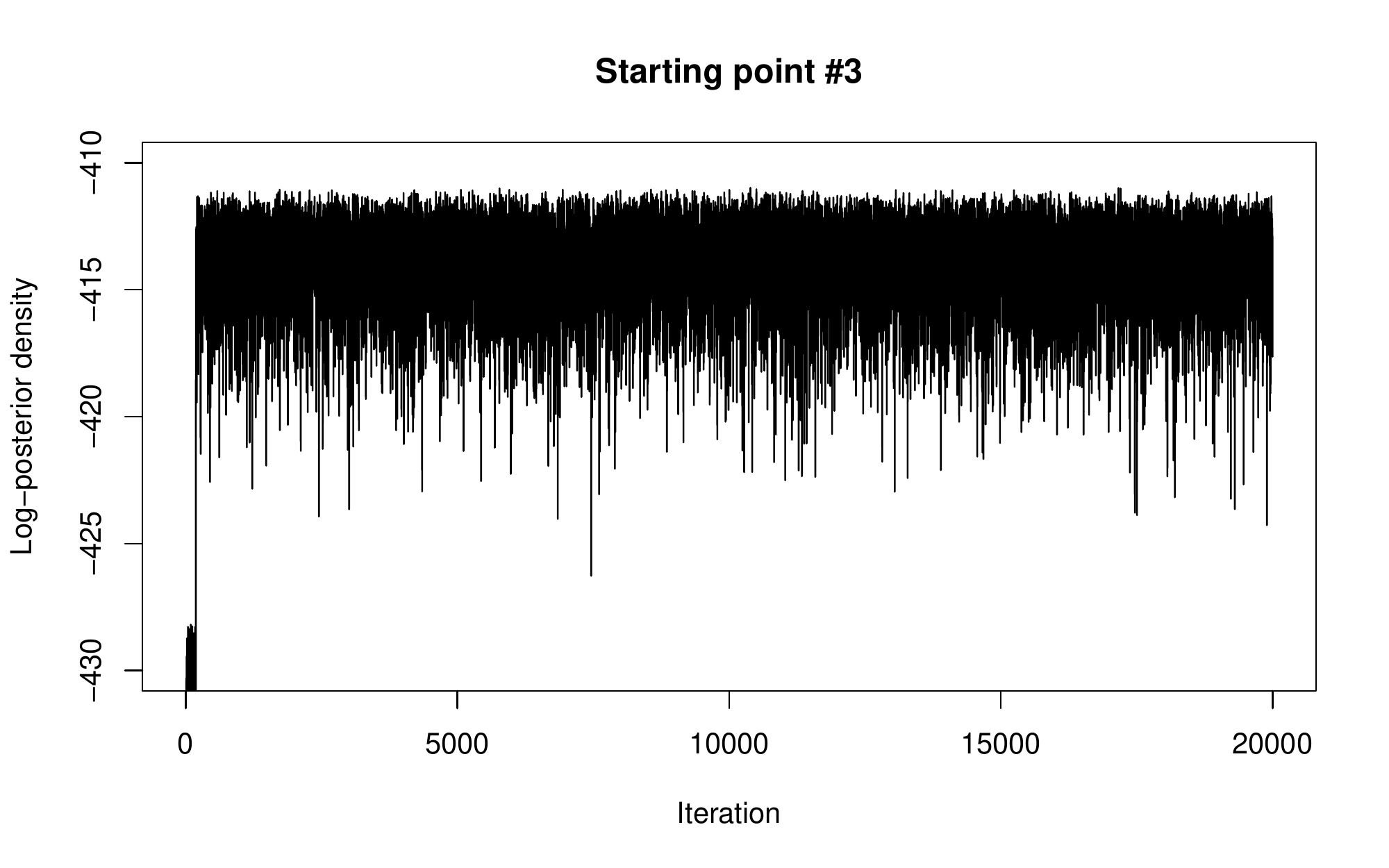}}
\subfloat{\includegraphics[scale=0.25]{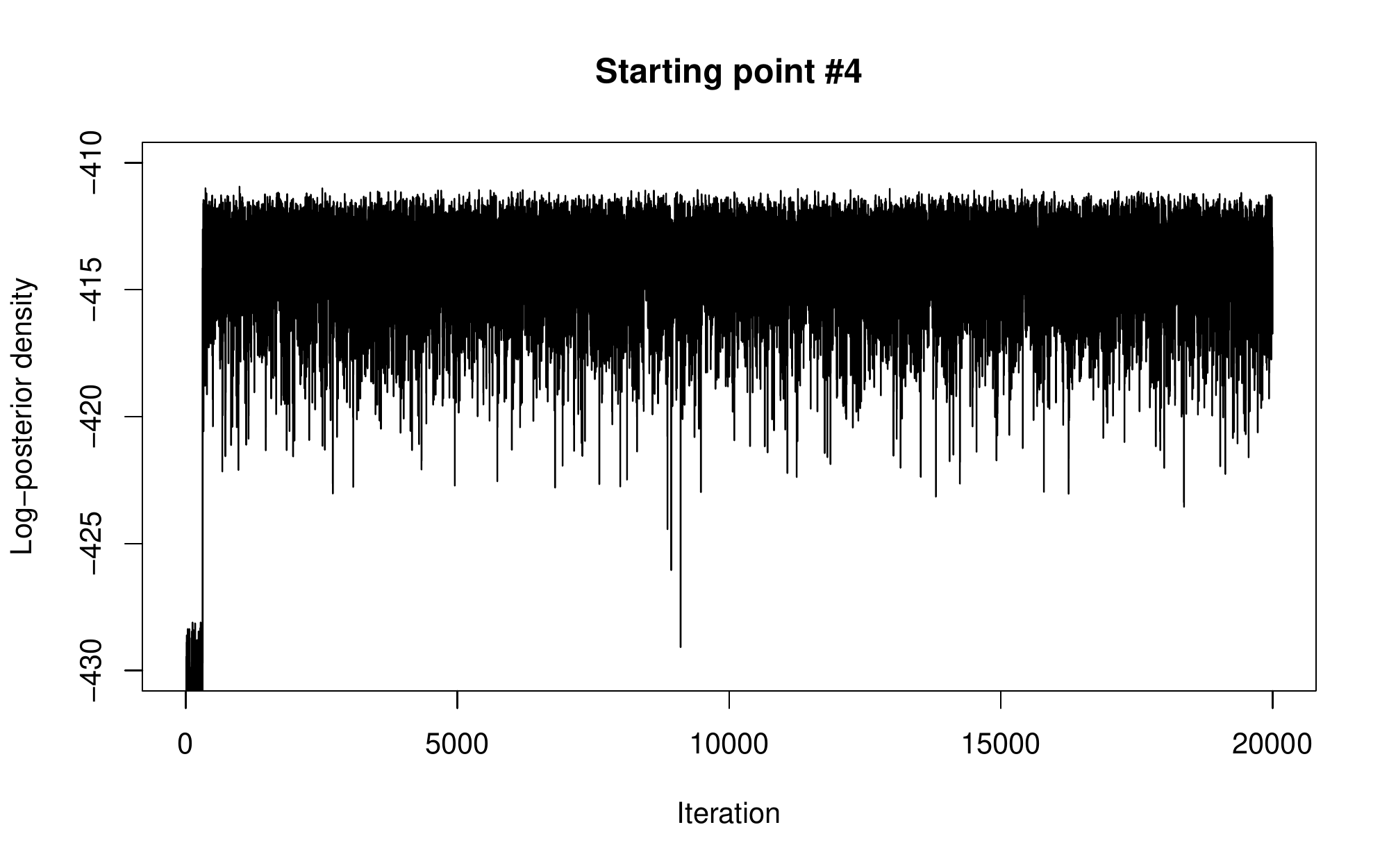}}\\
\subfloat{\includegraphics[scale=0.25]{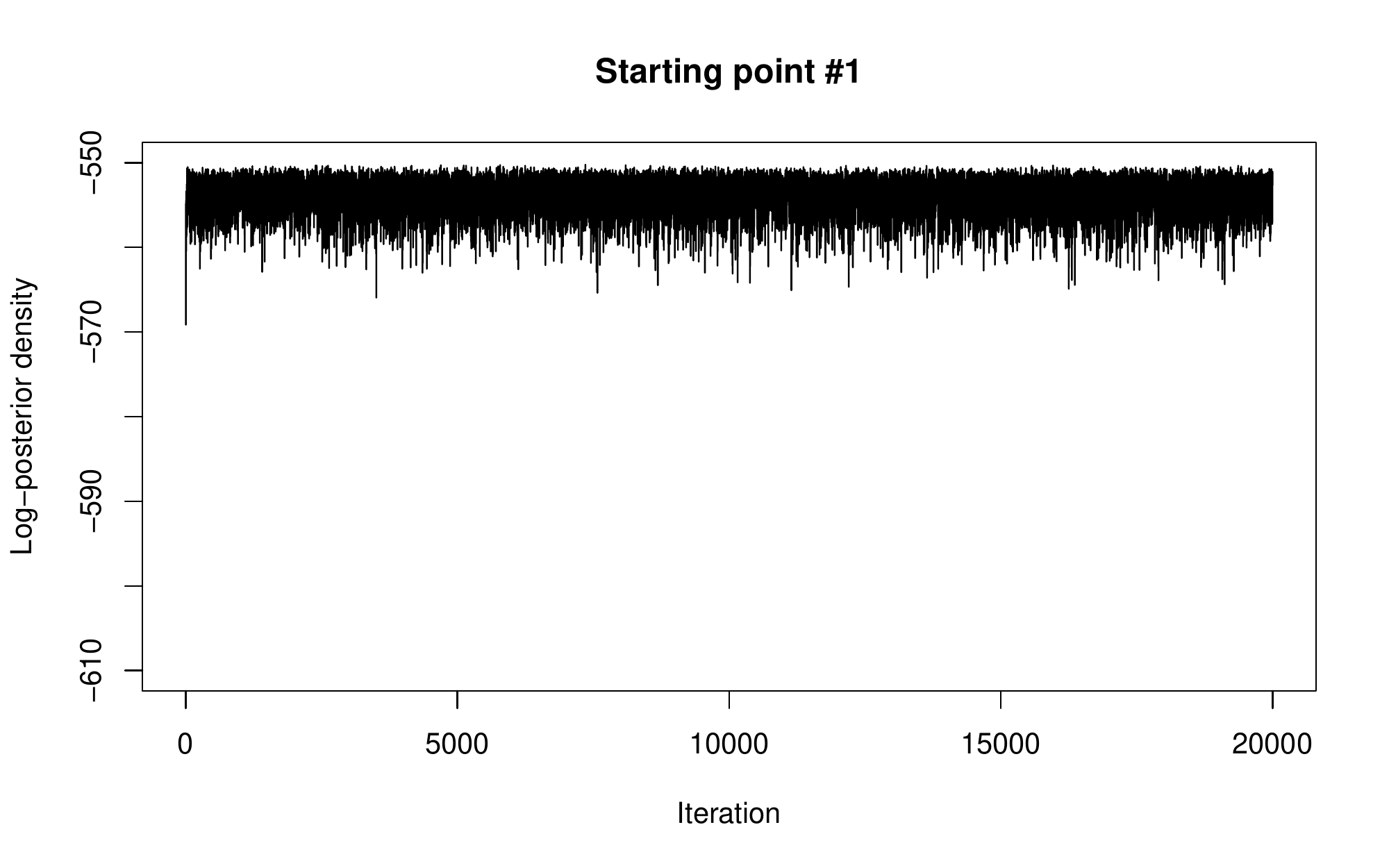}}
\subfloat{\includegraphics[scale=0.25]{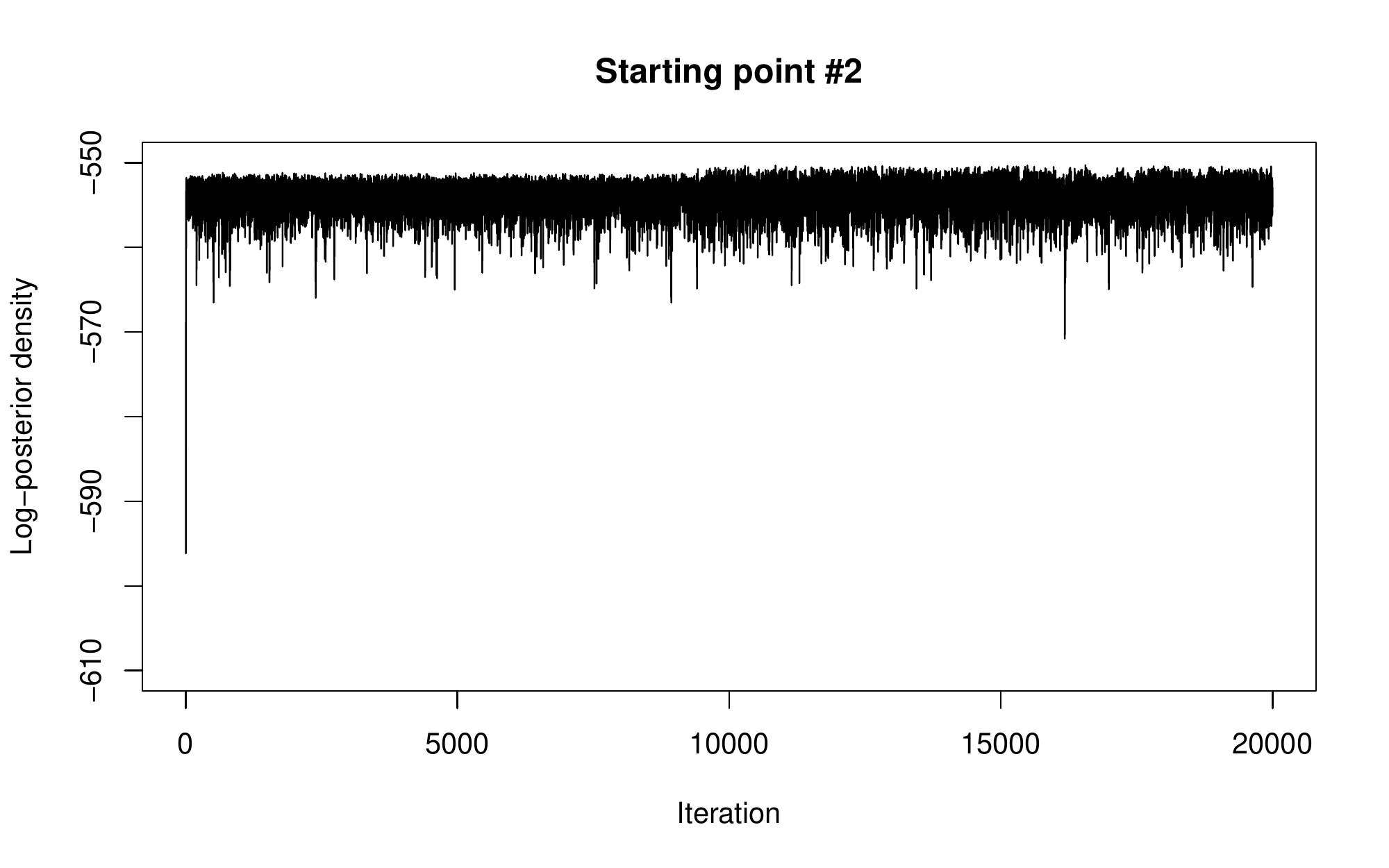}}
\subfloat{\includegraphics[scale=0.25]{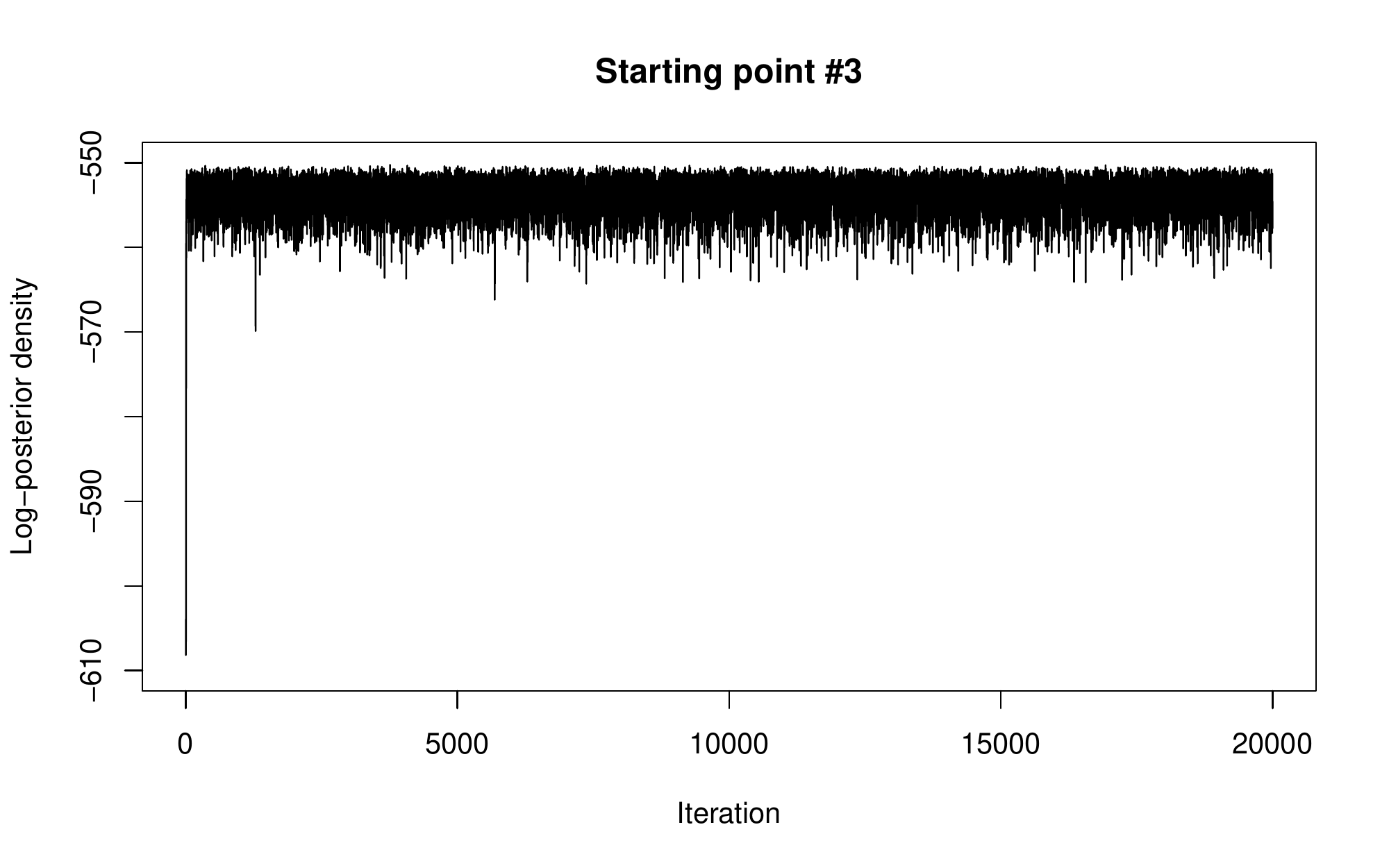}}
\subfloat{\includegraphics[scale=0.25]{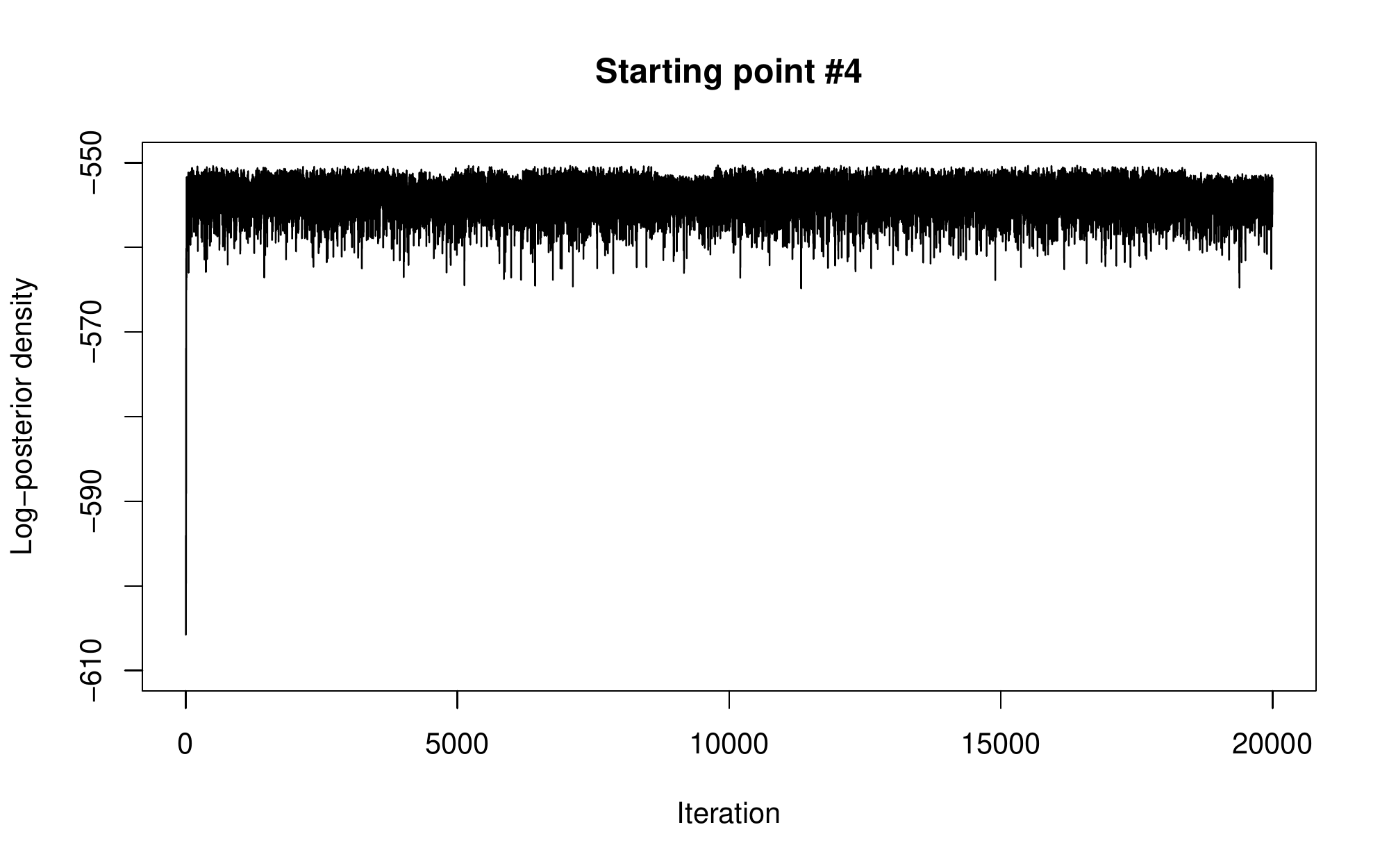}}
\caption{Traceplots of the MCMC chains launched with four random dispersed starting values for the two subsamples of the \texttt{sport} dataset: log-posterior\\ density of the subsample with $N_1=59$ units (upper panel) and log-posterior density of the subsample with $N_2=71$ units (lower panel).}
\label{fig:tracesSPORT}
\end{sidewaysfigure}

For the second real data application, we considered the \texttt{sport}
dataset included in \texttt{R} package \texttt{Rankcluster} \citep{Rankcluster}, where $N$=130 students at
the University of Illinois were asked to rank $K=7$ sports in order of
preference: $1$=Baseball, $2$=Football, $3$=Basketball, $4$=Tennis, $5$=Cycling, $6$=Swimming and 
$7$=Jogging. 
Prior to the Bayesian constrained EPL analysis, sample heterogeneity has been preliminary investigated
by using the EPL mixture methodology presented by \cite{Mollica:Tardella}, that suggested the presence of two preference groups with sizes $N_1=59$ and  $N_2=71$ (best fitting 2-component EPL mixture with BIC=2131.20).
We then decided to estimate the Bayesian constrained EPL separately on the two clusters, indicated with $\underline\pi_1^{-1}$ and $\underline\pi_2^{-1}$.
The posterior summaries of the EPL parameters are detailed in Tables \ref{t:agreeEPLsport} and \ref{t:agreeEPLsport2}, whereas Figure \ref{fig:tracesSPORT} shows the traceplots of the four MCMC chains launched with alternative starting values for the two subsamples. The modal reference order for each group turned out to be
$\hat\rho_1=(1,2,3,4,5,6,7)$ and $\hat\rho_2=(1,2,3,7,4,5,6)$, i.e., the forward order and a reference sequence with the top-3 sports as first choices followed by the final (7th) position assignment. The estimated group-specific modal orderings are
(7,6,4,5,3,1,2) and (1,2,3,4,6,7,5), indicating opposite preferences in the two subsamples towards team and individual sports.

\begin{table}[b]
\caption{Top-5 posterior probabilities of the reference order for the two subsamples of the \texttt{sport} dataset.}
\label{t:agreeEPLsport}
\centering
\subfloat[][Subsample 1] 
{\begin{tabular}{cc}
\hline
$\rho_1$ & $\pi(\rho_1|\underline\pi_1^{-1})$ \\
\hline
(1,2,3,4,5,6,7) & .9996\\
(1,7,2,3,6,4,5) & .0002\\
(1,2,3,4,7,5,6) & .0002\\
(1,2,7,3,6,4,5) & .0001\\
(1,2,3,7,6,4,5) & .0001\\
\hline
\end{tabular}\label{t:agreeEPL1sport}}\qquad\qquad
\subfloat[Subsample 2]
{\begin{tabular}{cc}
\hline
$\rho_2$ & $\pi(\rho_2|\underline\pi_2^{-1})$\\
\hline
(1,2,3,7,4,5,6) & .6643\\
(1,2,7,3,4,5,6) & .1195\\
(1,2,3,7,4,6,5) & .1083\\
(1,2,3,4,7,5,6) & .0761\\
(1,2,3,4,7,6,5) & .0171\\
\hline
\end{tabular}\label{t:agreeEPL2sport}}
\end{table}
\begin{table}[]
\caption{Posterior means of the support parameters for the two subsamples of the \texttt{sport} dataset.}
\label{t:agreeEPLsport2}
\centering
\begin{tabular}{rrr}
\hline
Sport & $\hat{p}_{1i}$ & $\hat{p}_{2i}$ \\
\hline
Baseball & .3883 & .0578 \\
Football & .2168 & .0515\\
Basketball & .1914 & .0834\\
Tennis & .0643 & .1565\\
Cycling & .0779 & .1579\\
Swimming & .0398 & .2122\\
Jogging & .0215 & .2807\\
\hline
\end{tabular}
\end{table}

\section{Conclusions}
\label{s:conc}

We have addressed some relevant issues in modelling and inferring choice behavior and preferences. 
The standard PL for complete  rankings relies on the hypothesis that the probability of a particular ordering does not depend on the subset of items from which one can choose (Luce's Axiom).
Although this can be considered a very strong assumption, the widespread use of the PL suggested to exploit it as a building block to
gain more flexibility. In particular, \cite{Mollica:Tardella} explored the possibility of adding flexibility with the help of two main ideas: i) the use of 
an additional discrete
parameter, the reference order, specifying the order of the ranks sequentially assigned by the individual and which should be inferred from the data; ii) the finite mixture of PL distributions enriched with the reference order parameter (EPL mixture). 

In this paper, we have focussed on i) and developed
a
methodology to infer the EPL distribution within the Bayesian framework, where additional monotonicity restrictions on  the reference order 
describe a 
``top-or-bottom'' attribution of the positions. After experiencing initial difficulties in implementing a well-mixing MCMC approximation, we have devised a hybrid strategy by combining appropriately tuned MH kernels and the GS. This allows for a successful exploration of the whole 
mixed-type 
parameter space. Compared with the previous frequentist approach, the resulting Bayesian inference turned out to be more efficient in achieving the most supported reference orders and provided a well-mixing MCMC simulation
which can be used to quantify the uncertainty on the EPL parameters. We stress that the previous attempts to infer on the underlying reference order were limited to the maximum likelihood point estimation via EM algorithm, which required a brute-force multiple starting point strategy in order to attempt to reach the global maximum. 
We note that,
with increasing $K$, 
 a fixed number of multiple starting points becomes a factorially decreasing (hence negligible) fraction of 
all
 the possible initializations
 and, obviously, this hampers the possibility of achieving eventually the global optimum in applications with larger $K$.
 On the other hand, 
 the sequential MCMC strategy provided valid results 
 which are 
 less sensitive to the starting reference order. 
 Hence, our Bayesian methodology provides 
both an appropriate way to assess parameter uncertainty
and  a computational improvement
for inferring the EPL.

Moreover, 
our model setup gains more insights on the sequential mechanism of formation of preferences
and whether it privileges a more or less naturally ordered assignment of the most extreme ranks. In other words, we show how it is possible to assess with a suitable statistical approach the formation of ranking preferences and answer through a statistical model the following questions: ``What do I start ranking first? The best or the worst? And what do I do then?''. 

Simulation studies confirmed the efficacy of the TJM-within-GS to recover the 
true 
EPL parameters generating the data, together with the benefits of the SM strategy to speed up the MCMC algorithm in the exploration of the posterior distribution. Moreover, the novel parametric approach was successfully applied to two real datasets concerning, respectively, a biomedical study and the analysis of preference patterns.


For possible future developments, several directions can be contemplated to further extend the Bayesian EPL with order constraints. First, the methodology can be generalized to infer on the unrestricted EPL, with the reference order taking values in the whole permutation set. Additional extensions could aim at the accommodation of partial orderings and at the introduction of item-specific and individual covariates
that can improve the characterization of the preference behavior. Moreover, a Bayesian EPL mixture could fruitfully support 
more efficiently
the identification of a 
parsimonious 
cluster structure in the
sample.

\appendix
\section*{Appendix}


Let us make an example in order to clarify the restrictions on the reference order space and the 
related notation. Let $\rho=(5,1,4,3,2)\in\tilde S_5$ be the reference order of interest. By comparing its entries with the forward  order $\rho_\text{F}=
(1,2,3,4,5)$ indicating the sequential assignment of top positions, the vector $\rho$ can be coded in binary format as follows
\begin{align*}
\text{Stage 1}\quad\rightarrow\quad F_1&=0,\qquad B_1=0,\qquad W_1=I_{[\rho(1)=\rho_\text{F}(1)]}=I_{[5=1]}=0,\\
\text{Stage 2}\quad\rightarrow\quad F_2&=0,\qquad B_2=1,\qquad W_2=I_{[\rho(2)=\rho_\text{F}(1)]}=I_{[1=1]}=1,\\
\text{Stage 3}\quad\rightarrow\quad F_3&=1,\qquad B_1=1,\qquad W_1=I_{[\rho(3)=\rho_\text{F}(2)]}=I_{[4=2]}=0,\\
\text{Stage 4}\quad\rightarrow\quad F_4&=1,\qquad B_1=2,\qquad W_1=I_{[\rho(4)=\rho_\text{F}(2)]}=I_{[3=2]}=0,\\
\text{Stage 5}\quad\rightarrow\quad F_5&=1,\qquad B_5=3,\qquad W_5=I_{[\rho(5)=\rho_\text{F}(2)]}=I_{[2=2]}=1,\\
\end{align*}
implying $\underline{F}=(0,0,1,1,1)$, $\underline{B}=(0,1,1,2,3)$ and $\underline{W}=(0,1,0,0,1)$. This means that, apart from the second and the fifth stages, the ranker always specifies her preferences by assigning bottom positions. With the help of both the forward and the backward order $\rho_\text{B}=(
5,4,3,2,1)$, 
we can expand the sequential formation of 
the inverse mapping 
in \eqref{ournotation}
from $\underline{W}$ to $\rho$ 
as follows
%
\begin{align*}
\text{Stage 1}\quad\rightarrow\quad \rho(1) & =\rho_\text{F}(1)^{0}\rho_\text{B}(1)^{1}=\rho_\text{B}(1)=5,\\
\text{Stage 2}\quad\rightarrow\quad \rho(2) & =\rho_\text{F}(1)^{1}\rho_\text{B}(2)^{0}=\rho_\text{F}(1)=1,\\
\text{Stage 3}\quad\rightarrow\quad \rho(3) & =\rho_\text{F}(2)^{0}\rho_\text{B}(2)^{1}=\rho_\text{B}(2)=4,\\
\text{Stage 4}\quad\rightarrow\quad \rho(4) & =\rho_\text{F}(2)^{0}\rho_\text{B}(3)^{1}=\rho_\text{B}(3)=3,\\
\text{Stage 5}\quad\rightarrow\quad \rho(5) & =\rho_\text{F}(2)^{1}\rho_\text{B}(4)^{0}=\rho_\text{F}(2)=2,
\end{align*}
yielding the initial sequence $\rho=(5,1,4,3,2)$.

\newpage

%

 \newcommand{\noop}[1]{}

\end{document}